# Not fit for Purpose: A critical analysis of the 'Five Safes'

Dr. Chris Culnane[1], Associate Professor Benjamin I. P. Rubinstein[1], Professor David Watts

**Abstract.** Adopted by government agencies in Australia, New Zealand and the UK as policy instrument or as embodied into legislation, the 'Five Safes' framework aims to manage risks of releasing data derived from personal information. Despite its popularity, the Five Safes has undergone little legal or technical critical analysis. We argue that the Fives Safes is fundamentally flawed: from being disconnected from existing legal protections and appropriation of notions of safety without providing any means to prefer strong technical measures, to viewing disclosure risk as static through time and not requiring repeat assessment. The Five Safes provides little confidence that resulting data sharing is performed using 'safety' best practice or for purposes in service of public interest.

## Introduction

Over the last few years the Five Safes framework has gained popularity in Australia, New Zealand and the UK as an approach to managing the risks of 'sharing' de-identified personal information. It claims to be a model whose basic premise is that personal information data sharing risks 'can be seen as a set of five 'risk (or access)'[2] dimensions: safe projects, safe people, safe data, safe settings, safe outputs.'[3]

The Five Safes has been adopted by a number of Australian governments and institutions as their preferred approach to managing the way they disclose (commonly called 'share') de-identified personal information. The list includes the Australian Bureau of Statistics,[4] the Office of the National Data Commissioner,[5] the Australian Institute of Health and Welfare,[6] the Commonwealth Scientific Industrial and Research Organisation,[7] Data.NSW[8] and the Victorian Government.[9]

In these instances, the Five Safes functions as a policy instrument. It does not have the force of law and is thus subordinate to it. In practice the relevant applicable laws are information privacy laws.

That said, the Five Safes has also been embodied in South Australian legislation, the *Public Sector (Data Sharing) Act 2016*.[10] South Australia is one of only two Australian jurisdictions without a public sector information privacy law.

Australia's Office of the National Data Commissioner, currently a business unit within the Commonwealth Department of Prime Minister and Cabinet, has developed an exposure draft of legislation, called the *Data Availability and Transparency Bill* (formerly called *Data Sharing and Release*[11]), which was released for comment

---

[1] School of Computing and Information Systems, The University of Melbourne, Australia.
[2] The reference to 'access' is a first example of the problematic use of terminology in this area of scholarship, in particular the disjunction that appears to exist between data scientists, lawyers and policy makers. Australia's information privacy laws give individuals rights to control the collection, use and disclosure of their personal information. Where an organisation permits a third party to 'access' an individual's personal information it has *disclosed* that information to the third party.
[3] Tanvi Desai, Felix Ritchie, Richard Welpton, *Five Safes: designing data access for research,* University of the West of England Bristol, Economics Working Paper Series 1601, p5. See https://www2.uwe.ac.uk/faculties/BBS/Documents/1601.pdf (January 2020)
[4] Australian Bureau of Statistics, *Managing the Risk of Disclosure: The Five Safes Framework,* 1160.0 – ABS Confidentiality Series, August 2017. See https://www.abs.gov.au/ausstats/abs@.nsf/Latestproducts/1160.0Main%20Features4Aug%202017 (January 2020)
[5] Office of the National Data Commissioner, *Data Sharing Principles*, March 2019. See https://www.pmc.gov.au/resource-centre/public-data/data-sharing-principles (January 2020)
[6] Australian Institute of Health and Welfare, *The Five Safes Framework*. See https://www.aihw.gov.au/about-our-data/data-governance/the-five-safes-framework (January 2020)
[7] Christine O'Keefe, Stephanie Otorepec, Mark Elliot, Elaine Mackey, and Kieron O'Hara, Commonwealth Scientific and Industrial Research Organisation, Data 61 and the Office of the Australian Information Commissioner, *The De-identification Decision-Making Framework,* CSIRO Reports EP173122 and EP175702, 2017, in particular chapter 3. See https://publications.csiro.au/rpr/download?pid=csiro:EP173122&dsid=DS3 (January 2020)
[8] Data.NSW, *Data Sharing Principles,* 19 June 2019. See https://data.nsw.gov.au/data-sharing-principles (January 2020)
[9] Victorian Government, *Data Legislation, Security and Privacy,* Assessing and Mitigating Risk, We use a trusted data access and sharing model. See https://www.vic.gov.au/data-security-privacy (January 2020)
[10] See s7 *Public Sector (Data Sharing) Act 2016* (SA)
[11] Australian Government, Department of Prime Minister and Cabinet, *Data Sharing and Release Legislative Reforms Discussion Paper,* September 2019, p53. See https://www.datacommissioner.gov.au/sites/default/files/2019-09/Data%20Sharing%20and%20Release%20Legislative%20Reforms%20Discussion%20Paper%20-%20Accessibility.pdf (January 2020)

September 14 2020.[12] The *Data Sharing and Release* Discussion Paper foreshadowed that the legislation 'will provide *legal* grounds to empower the government to share public sector data for specified purposes with the right to safeguards.'[14] The safeguards it has proposed are built on the *Data Sharing Principles*.[15] The Data Sharing Principles are the Five Safes.[16] The Five Safes are thus in the process of transitioning from policy to law, at least within the Commonwealth public sector.

Despite its significant adoption in Australia, New Zealand and the UK[17], the Five Safes has received little, if any, critical attention. This short paper aims to remedy this. We argue that the Five Safes is problematic for a number of reasons. It is disconnected from the legal protections conferred under information privacy law and thus fuels misconceptions about individual rights to exercise a degree of control over the disclosure of personal and sensitive information. It appropriates notions of safety without being anchored in any objective standard by which to assess or measure what is and is not safe. Instead of clarifying existing legal concepts that govern the use and disclosure of personal and sensitive information it overlays these with yet another series of ambiguous tests based on safety assessments. It fails to account for the dynamic nature of risk because it focuses on risk at a single point in time rather than assessing it on a whole of information lifecycle and end-to-end basis. It encourages a 'tick the box' approach to risk assessment and mitigation, without any preference for strong technical measures such as cryptography or differential privacy. It privileges the interests of governments and their institutions over those of individual citizens because it aims to 'unlock the potential of..data'[18] in a manner that does not necessarily serve the public interest.

## Open government data

Over the last decade governments across the world have become increasingly preoccupied with 'opening up' the data they collect and hold to enable its use and re-use. This phenomenon is part of the open government data movement.

Open government data is a policy approach to government information that seeks to promote 'transparency, accountability and value creation by making government data available to all.'[19] Open government data is said to create other benefits such as enabling better public participation and collaboration in the creation of innovative, value added services; to improve the public's ability to make decisions and to inform choices; better relations between government and citizens; to improve evidence-based decision-making and as a source of economic growth, social innovation and new forms of entrepreneurship.[20]

Initially, the data covered by open government data initiatives included categories such as business information, government registers, geographic information, legal information, meteorological information, social data and transport information. However, as open government data policies have evolved it has come to encompass data that is, or has been derived from, personal information: 'government data' includes personal information collected by government in undertaking its functions.

---

[12] Office of the National Data Commissioner, Australia, *Consultation on the Data Availability and Transparency Bill 2020*, 14 September 2020. See https://www.datacommissioner.gov.au/exposure-draft (October 2020)
[14] *Ibid,* p3. Our emphasis.
[15] Australian Government, Department of Prime Minister and Cabinet, *Better Practice Guide to Applying Data Sharing Principles,* Version 15 March 2019, p6. See https://www.pmc.gov.au/sites/default/files/publications/data-sharing-principles-best-practice-guide-15-mar-2019_0.pdf (January 2020)
[16] *Ibid,* Chapter 3
[17] Australian Bureau of Statistics, *Managing the Risk of Disclosure: The Five Safes Framework,* 1160.0 – ABS Confidentiality Series, August 2017. See https://www.abs.gov.au/ausstats/abs@.nsf/Latestproducts/1160.0Main%20Features4Aug%202017 (January 2020)
[18] Australian Government, Department of Prime Minister and Cabinet, *Data Sharing and Release Legislative Reforms Discussion Paper,* September 2019, p13. See https://www.datacommissioner.gov.au/sites/default/files/2019-09/Data%20Sharing%20and%20Release%20Legislative%20Reforms%20Discussion%20Paper%20-%20Accessibility.pdf (January 2020)
[19] Organisation for Economic Cooperation and Development, *Open Government Data.* See https://www.oecd.org/gov/digital-government/open-government-data.htm (December 2019)
[20] Barbara Ubaldi, OECD Working Papers on Public Governance No.22, *Open government Data: Towards Empirical Analysis of Open Government Data Initiatives,* 27 May 2013. See https://www.oecd-ilibrary.org/governance/open-government-data_5k46bj4f03s7-en (December 2019)

## Information privacy and open government data

In Australia, personal information is protected by Commonwealth, State and Territory information privacy laws.[21] The Commonwealth *Privacy Act 1988* regulates the collection and handling of personal and sensitive information in the Commonwealth public sector and in the private sector on a national basis. State and Territory information privacy laws apply to the collection and handling of personal and sensitive information in their respective public sectors. In aggregate, these pieces of legislation produce what is widely considered to be a 'patchwork' of information privacy protections. Although they are implemented differently, they are each built around a common set of privacy principles that govern the collection, use, disclosure and other handling of personal and sensitive information.[22] Despite the patchwork, they share common concepts and definitions.

One of the key misconceptions about information privacy law is that it prevents appropriate use and disclosure of personal and sensitive information. This is incorrect. Personal information can *always* be shared with consent. Moreover, privacy is not an absolute right. Outside of consent, information privacy law *has always* permitted information to be shared where there is a recognised, countervailing public interest. For example, personal information can be used and disclosed for a purpose – a secondary purpose – related to the purpose for which it was collected[23] or where there is legal authority to do so.[24] Collection and disclosure of health information for the purposes of public health and safety research is also permitted under this type of framework.[25]

In public policy terms, a Five Safes use and disclosure framework that displaces these requirements is a significant dilution of privacy rights. Where it is used in parallel to existing safeguards it superimposes a risk framework on them which, as we discuss below, is not fit for purpose.

## Personal information and de-identification

Personal and sensitive information collected by government, as a sub-category of 'government data,' cannot be published as part of open government data initiatives because information privacy laws prohibit it. This has meant that open data advocates have sought a means to leverage government-held personal and sensitive data without breaching information privacy law. The approach they have settled on is de-identification.

The *Privacy Act 1988* defines 'personal information' as:

> information or an opinion about an identified individual, or an individual *who is reasonably identifiable:*
>
> (a) whether the information or opinion is true or not; and
> (b) whether the information or opinion is recorded in a material form or not.[26]

One of the key interpretative issues associated with the definition is the meaning of 'reasonably identifiable.' Some assistance is provided by the definition of 'de-identified:'

> de-identified: personal information is **de-identified** if the information is no longer about an identifiable individual or an individual who is *reasonably identifiable.*[27]

The combined effect of these definitions is that where information is about an individual who is identified or reasonably identifiable, that information is regulated by information privacy legislation. The corollary is that where information is not about an individual who is identified or who is reasonably identifiable, the information is de-identified, is not regulated by information privacy law and can be used and disclosed (i.e., shared) without privacy restrictions.

---

[21] These are the *Privacy Act 1988* (Cth), the *Privacy and Personal Information Protection Act 1998* (NSW), the *Privacy and Data Protection Act 2014* (Vic), the *Information Privacy Act 2009* (Qld), the *Personal Information Protection Act* (Tas), the *Information Privacy Act 2014* (ACT), the *Information Act 2002* (NT). There is no public sector information privacy legislation in either South Australia or Western Australia. Some jurisdictions have enacted specific legislation that separately covers health information privacy. These are the *Health Records and Information Privacy Act 2002* (NSW), the *Health Records Act 2002* (Vic) and the *Health Records (Privacy and Access) Act 1997* (ACT).

[22] These are the OECD's *Guidelines on the Protection of Privacy and Transborder Flows of Personal Data.* See https://www.oecd.org/internet/ieconomy/oecdguidelinesontheprotectionofprivacyandtransborderflowsofpersonaldata.htm (January 2020)

[23] Australian Privacy Principle (APP) 6.2(a)(i). Where the personal information is also sensitive information the purpose must be *directly* related. See APP 6.2(a)(ii)

[24] APP 6.2(b).

[25] For example, under s16B of the *Privacy Act 1988.*

[26] Our emphasis. See s6(1) *Privacy Act 1988* (Cth). Equivalent State and Territory definitions are substantially similar.

[27] Our emphasis is to the words in italics. The word 'de-identified' is in bold in the definition. See s6(1) *Privacy Act 1988* (Cth).

The problem with these legal tests is that they provide little guidance about what needs to be done to personal information to transform it from being identifiable or reasonably identifiable to being de-identified.

De-identification is one of the most hotly contested and controversial issues in international data protection law and policy. Sometimes these are framed against several similar (but not necessarily identical) terms such as 'anonymisation,'[28] pseudonymisation,'[29] 'redaction,' and 'confidentialisation.'[30] None of these terms are used in Australian information privacy law and their use is more likely than not to add complexity to the legal and policy landscape.

The US National Institute of Science and Technology (NIST) states that:

> [d]e-identification is a tool that organizations can use to remove personal information from data they collect, use, archive, and share with other organizations..(it) is not a single technique, but a collection of approaches, algorithms, and tools that can be applied to different kinds of data with different levels of effectiveness. In general, privacy protection improves as more aggressive de-identification techniques are employed, but less utility remains in the resulting dataset.[31]

There are a variety of techniques that can be used to de-identify personal information. These include sampling, removing quasi-identifiers, data perturbation, rounding and using encrypted identifiers, to name but a few. Frameworks with mathematical definitions of de-identification have been developed including k-anonymity[32]. None have proven to reliably prevent high utility de-identified unit record level data from being re-identified. Differential privacy[33] provides strong guarantees against re-identification, typically when releasing aggregate data. It has been adopted by the U.S. Census Bureau for all releases from the 2020 U.S. Decennial Census.[34]

The Five Safes is intended to address the problems inherent with "de-identification", without prescribing any technical measures, definitions of de-identification, or definitions of "reasonable" identifiability.

## The notion of safety

At the core of the Five Safes is the notion that a safety can be achieved across some or all of its modalities. This is our threshold point of contention. Safety it is not an absolute position. It is a position on an unspecified and undefined risk continuum. Although it may be theoretically possible to achieve a "safe" state, this can only be achieved against absolute positions that would render the undertaking pointless.

For example, "safe" people could be achieved by denying anyone access. Likewise, "safe" data could be achieved by deleting all the data. If we assume that such absolutes are unrealistic, and that someone will have access to something, we can be certain that rather than there being an absolute "safe" state, the posture is in fact a point on a scale of safety. Even if such a scale were to exist, it is unrealistic for it to be applied consistently and measurably across the variety of organisations that seek to rely on it.

---

[28] Anonymisation is used in the European Union's *General Data Protection Regulation* to signify 'information which does not relate to an identified or identifiable natural person or to personal data rendered anonymous in such manner that the data subject is not or no longer identifiable.' See Recital 26, GDPR.
[29] Pseudonymisation is also a term that is used in the GDPR. It 'means the processing of personal data in such a manner that the personal data can no longer be attributed to a specific data subject without the use of additional information, provided that such additional information is kept separately and is subject to technical and organisational measures to ensure that the personal data are not attributed to an identified or identifiable natural person.' See Article 4(5) GDPR.
[30] 'Confidentialisation' is a term used by the Australian Bureau of Statistics and is referred to in its Confidentiality Information Series which 'explains how to assess and manage confidentiality risks by using tools such as the Five Safes Framework, and outlines methods for treating data as part of this risk management approach.' See
https://www.abs.gov.au/AUSSTATS/abs@.nsf/Latestproducts/1160.0Main%20Features1Aug%202017?opendocument&tabname=Summary&prodno=1160.0&issue=Aug%202017&num=&view= (January 2020)
[31] Simson Garfinkel, National Institute of Science and Technology, *De-identification of Personal Information,* NISTIR 8053, October 2015, p1. See https://nvlpubs.nist.gov/nistpubs/ir/2015/NIST.IR.8053.pdf (January 2020)
[32] Pierangela Samarati and Latanya Sweeney, *Protecting privacy when disclosing information: k-anonymity and its enforcement through generalization and suppression*, technical report, SRI International, May 1998
[33] Cynthia Dwork, Frank McSherry, Kobbi Nissim and Adam Smith, *Calibrating noise to sensitivity in private data analysis*, in Theory of Cryptography Conference, pp. 265-285, March 2006
[34] John M. Abowd, *The U.S. Census Bureau Adopts Differential Privacy*, Proceedings of the 24th ACM SIGKDD International Conference on Knowledge & Data Mining, July 2018

## Safe People

The notion that it is possible to define safe people is flawed. Defence, intelligence, national security and law enforcement agencies dedicate significant resources to vet their employees so that they can safely be entrusted with security classified data. Despite this, there are numerous examples where those security vetted individuals betray their employers' trust.

To further complicate the evaluation, the notion of "safe" people is not static. It is inherently dynamic because it depends on the motivations of the individual, which are ever changing. It is also important to recognise that an unsafe person is not necessarily a sophisticated adversary, or even a person intent on causing harm. In many cases it is little more than curiosity that causes a breach of trust, such as searching for information about a neighbour, friend, or partner. The motivation to attempt such a query is intrinsically linked to the likelihood of getting caught. If there is only a minimal chance of being discovered the temptation will be strong, the payoff is large and the risk is close to zero.

The evaluation of "safe" people is dependent on safe settings, i.e., those settings which involve regular reviews and audits of the queries run on a dataset are likely to produce safer people because the likelihood of being caught out is high. Contrary to its claims, the framework's dimensions of safety are not necessarily independent.

## Safe Data

The concept of "safe" data is a misnomer. What does it mean to describe data as safe? If data was safe in absolute terms, what is the relevance of the remaining four safes? As noted earlier, information privacy law takes the approach that personal data is safe to share if it has been de-identified but not otherwise.

There is a long history of the failure of technical de-identification measures,[35] [36] often in publicly released data sets, the release of which can harm those who were re-identified.[37] Awareness of the failure of de-identification is not new, Paul Ohm famously described the fallacy of de-identification in 2010.[38] Yet organisations and policy makers have failed to heed the warning, and we continue to see ill-defined de-identification advocated as a method for creating "safe" data. In some situations, differential privacy or cryptography offer rigorous measures of safe data, however the Five Safes framework offer no means by which to prefer such measures. An organisation, having adopted Five Safes, can declare data to be safe without there being any safety from re-identification (for example) at all.

## Safe Projects

An evaluation of the safety of a project assumes that the intention of the project's initiator can be accurately determined, and that all future potential uses of the data are predictable and are known or knowable at the inception of an information sharing project. This is a questionable assumption. It ignores the fact that information-based projects frequently feed into other information projects, each building on each other and each having different objectives.

Safe Projects depend on Safe People, since the project description will be self-reported. The counter to this is that Safe Projects ensure that data will only be used for appropriate public interest research, and not, for example, commercial gain or in processes that could cause harm to the individuals within the dataset. However, such a counter point assumes that there are Safe People who would be willing to propose unsafe projects. If such a person was to exist, how could they continue to be considered a Safe Person? Evaluations of projects should take place as minimum

---

[35] Chris Culnane, Benjamin I. P. Rubinstein and Vanessa Teague, *Health Data in an Open World*, arXiv:1712.05627, Dec 2017
[36] Chris Culnane, Benjamin I. P. Rubinstein and Vanessa Teague, *Stop the Open Data Bus, We Want to Get Off*, arXiv:1908.05004, Aug 2019
[37] Arvind Narayanan and Vitaly Shmatikov, *Robust de-anonymization of large sparse datasets*, Proceedings of the 2008 IEEE Symposium on Security and Privacy, pp. 111-125, 2008
[38] Paul Ohm, *Broken promises of privacy: Responding to the surprising failure of anonymization*, UCLA Law Review, Vol. 57, 2010. See http://paulohm.com/classes/techpriv13/reading/wednesday/OhmBrokenPromisesofPrivacy.pdf (January 2020)

due diligence, but only offer value if reassessment is ongoing, and is accompanied by serious consequences for failure to meet guarantees of maintaining safe project status.

### Safe Environment

Notions of safer environments can have merit, but truly protective environments are challenging to setup while Five Safes offers no specific guidance whatsoever. A secure research environment, which is both physically and technologically secure can significantly increase levels of data protection. Such facilities exist in a number of organisations, including the UK Office for National Statistics. However, such environments are dependent on continuous auditing and statistical disclosure measures being applied to results that are to leave the environment. Such measures are costly and are inherently unscalable. Remote secure environments offer far less protection but are sometimes viewed as equivalent. In reality, a remote secure environment is only secure if the user chooses to abide by the security mechanisms and does not attempt to circumvent them. For example, it would be easy to extract data from a secure remote environment by recording screen shots and using automated data extraction tools.

Where secure remote environments are beneficial is in reducing the risk of outright loss of the data file. If the file does not reside on the machine it is not possible to lose it due to machine theft or targeted attack. However, if an adversary is sufficiently motivated it will be able to utilise the screen scraping approach to extract the data - although such an attack would be costly and risky to undertake.

Overall, we can see that anything but offline secure research environments are inherently dependent on safe people. If the people turn out to not be safe, then a remote safe environment will not be sufficient to protect the data.

### Safe Outputs

De-identification is in general an unreliable approach to safe outputs; were it successful, the full suite of Five Safes may not be required. While differential privacy can in certain cases *prevent* re-identification while maintaining utility of releases, Five Safes does not favour one technical measure above any other e.g., once popular k-anonymity which lacked any attacker model or security property and has succumbed to attack.[39]

Furthermore, unless the safe output process is conducted by the owning authority, verification of safe output can be prohibitively challenging. In many cases, Safe Outputs are dependent on both Safe People, and Safe Environments.

## Safety vs. risk

The Five Safes framework is based on the concept of evaluating safety, instead of risk. We argue that such an approach is fundamentally flawed and inconsistent with best practice in the wider security field. It is a well-known trope that there is no such thing as a 100% secure system. There are always vulnerabilities and risks. The objective of building secure systems is to be able to quantify and subsequently minimise the risk that one of those vulnerabilities could be exploited. For example, consider cryptography key sizes, where evaluation is based on the risk of attack, and the capability of the attacker. A particular key size is considered secure up until a certain date, against a certain level of attacker. The specification lists the point where it breaks, not the attackers it is secure against.

### Minimise risk or increase safety

The issue with "safety" is not just a matter of terminology: it is essential to maintain a mindset to consider the attacker, and evaluate the risk of such an attacker succeeding, instead of striving for an arbitrary safety classification. Consider a cyclist setting out on a bike ride. A safety focused approach dictates wearing a helmet, a high visibility jacket, having a well-maintained bike, with lights and a bell, and verification that the cyclist is proficient in cycling. However, the greatest risk to the cyclist is a third-party motorised vehicle. The safety measures will undoubtedly help reduce risks, and should the rider be cycling in isolation they could be described as rendering the activity safe. However, when in an uncontrolled environment, with competing third party interests, such safety measures will have almost no effect in incidents with distracted, speeding, drunk, or unqualified drivers. It would be far more effective to evaluate whether to undertake the bike ride based on the weather, the time of day, the route, and traffic. Such an evaluation is inherently dynamic, and is risk focused, it accepts that the base safety measures will already be taken, whilst simultaneously considering the external risks. In the world of data privacy, linkable auxiliary datasets take the role of the third-party driver, often with the same non-trivial risk that can easily go unaccounted.

---

[39] Ashwin Machanavajjhala, Daniel Kifer, Johannes Gehrke and Muthuramakrishnan Venkitasubramaniam, *L-diversity: Privacy beyond k-anonymity*. ACM Transactions on Knowledge Discovery from Data, 1(1), 2007.

Importantly when evaluating risk instead of safety, there is an implicit acceptance that there is no such thing as a risk-free activity. The opposite is not true when evaluating safety to achieve a state of "safe". The very wording within the Five Safes creates the impression that an absolute state of "safe" can be achieved, when in fact it never can. Furthermore, by utilising an emotive word like "safe" it disproportionately creates an overly optimistic sense of well-being where none may exist. An activity described as "safe" seems far more approachable than the same activity described as "low risk".

## Five Safes collectivity

One of the cornerstones of the Five Safes framework is that different safes can be used to compensate for a lack of safety in one with the other. This has been described as independence of safety dimensions[40] and as a graphics equaliser,[41] in which values can be increased to create a balanced sound. However, we would argue that such a view is flawed. Rather than the safes being complementary, we would argue that several are dependent, and rather than being able to counter weakness in one with strength in another – a would-be instance of "defence in depth" that is an important paradigm in computer security – often weakness in one will lead to compromising the strength of another. By way of an example, we will analyse the situation where a person is not vetted to a high level and therefore is not considered to be highly safe. We will analyse whether it is reasonable to counter such a weakness with additional safety in the other four safes.

### Increasing Safe Environment

As discussed the safety of a person impacts on the safety levels of everything else. Even if the safest environments are used, in the form of secure research environments, such environments, by virtue of being safe environments only permit safe people in, often vetted to a much higher level than anything else. It is difficult to imagine a scenario where a person not considered highly safe would be permitted into a secure research environment.

Consider the secure *remote* environment. As discussed, safety of such environments relies on the user not actively trying to circumvent the protections. Such environments can be subverted by either a direct adversary (an unsafe person), or a third party through compromise of the user's machine. Even in the case of secure research environments, a highly motivated attacker would be able to extract a limited amount of information, generally bounded by the amount of information they can remember. This is why protections within safe environments must be close to what is required externally. Real examples of such a weakness have been shown[42] in which it would be feasible to target an individual using nothing more than what could be remembered by an analyst and typical tools available in secure environments.

### Increasing Safe Projects

The first challenge to increasing the level of project safety is the assumption of independence with trust in the accessing party's intentions – a safe person assumption. A malicious user could provide a description of a safe project and then subsequently use the data for alternative purposes. Banning users does not reduce harm already caused.

Few research projects follow plans exactly. Consider the provision of a commercial dataset. If a research team were to acquire a dataset within the scope of a particular project, there could be motivation for a third-party company, without direct access, to collaborate with the researchers to run analysis or link the sensitive data with their own.

---

[40] Australian Bureau of Statistics, *Managing the Risk of Disclosure: The Five Safes Framework,* 1160.0 – ABS Confidentiality Series, August 2017. See https://www.abs.gov.au/ausstats/abs@.nsf/Latestproducts/1160.0Main%20Features4Aug%202017 (January 2020)
[41] Steve McEachern, *Implementation of the Trusted Access Model*, ASSA Policy Roundtable, Australian Data Archive, Nov 2015. See http://rsss.anu.edu.au/sites/default/files/SMcEachern_researcherperspective_ASSANov2015.pdf (January 2020)
[42] Chris Culnane, Benjamin I. P. Rubinstein and Vanessa Teague, *Vulnerabilities in the use of similarity tables in combination with pseudonymisation to preserve data privacy in the UK Office for National Statistics' Privacy-Preserving Record Linkage*, arXiv:1712.00871, 2017.

### Increasing Safe Data

There are numerous examples of data being released that was believed to be safe which subsequently turned out not to be the case. The Five Safes framework itself does nothing to avoid this situation. Most data protection schemes, for example k-anonymity, are partially reliant on safe people and are rarely tested against active adversaries; many schemes are dependent on accurate knowledge of data sources in a potential attacker's possession. Such assumptions may *lower* bound the risk of re-identification, only. Where robust data protection methodologies are used, for example, differential privacy in releasing aggregate statistics to a third parties, the need for the Five Safes is negated. The data curator following Five Safes might increase perturbation, suppression, and aggregation to obtain an impression of increasing privacy protection, while in reality they only reduce utility with incremental and hard-to-quantify privacy protection.

### Increasing Safe Outputs

In all but the case of the highest level of Safe Environment, the output will be controlled and produced by an end user or curator. As such, the quality of the application of Safe Outputs depends on Safe People. Conversely, requiring a high degree of Safe Output may not prevent a non-compliant, high risk person from releasing sensitive data. Finally, in other cases such as in the application of differential privacy, it is difficult to distinguish Safe Data and Safe Outputs, making the caveats for the former relevant here.

### Less graphics equaliser, more house of cards

A better analogy for the Five Safes framework would be a house cards, rather than a graphics equaliser. It is not that different safes can balance each other, it is that if one safe fails it can bring down all or some of the rest. Five Safes discourages security best practice such as defence in depth. Clearly not all safes are equal, people being one of the most critical, whilst projects are important, but appear difficult to operationalise. It is difficult to see how a failure in Safe Projects could undermine Safe People. However, as we have argued, it is clear that a failure in evaluating Safe People could undermine all other safes.

The exact nature of the failure of security clearances is beyond the scope of this paper, but there have been sufficiently many to indicate the process is fundamentally flawed and cannot be wholly relied upon. The framework does nothing to indicate how to determine Safe People or other dimensions of safety, leaving it up to organisations to determine and self-regulate even though they may have no experience in conducting such a programme, or have pressures that are in tension with best practice data protection measures.

## Conclusion

In this paper we have provided a holistic analysis of the Five Safes model and described why it is not suitable as a model for determining the safe release of data. Unfortunately, despite the failings within the framework it continues to be advocated as an approach. It is even being expanded to include more safes.[43] It is far from clear how adding additional levels to the house cards will produce a more stable and successful framework. The structural dependencies will remain, but even greater false confidence will ensue.

The sharing of data is crucial to science, medical research and evidence-based policy making, but it can come at an undue cost to individual privacy. Any approach to managing disclosure risk should include the level of rigour seen in security and cryptography, rather than ad-hoc models that increase data sharing without providing any specific security properties.

False steps in much needed privacy legislative and policy updates, could ultimately erode trust in data collection, and encourage the public to consider counter measures. I.e., by providing false data. Such contamination of datasets will have a damaging effect on both research and policy and undermine the very conclusions and developments that motivate complete data collections. Legislators must address these concerns before pushing ahead with greater sharing, and data end users must demonstrate a greater appreciation for the responsibility they have to protect the privacy and safety of individuals.

---

[43] Australian Computer Society, *Privacy-Preserving Data Sharing Frameworks: People, Projects, Data and Output*, Dec 2019.